*(abstract of the report at ESR2007 conference)*

# Polynomially scaling spin dynamics simulation algorithm based on adaptive state space restriction

Ilya Kuprov, Nicola Wagner-Rundell, P.J. Hore

*Department of Chemistry, University of Oxford,*
*Physical and Theoretical Chemistry Laboratory,*
*South Parks Road, Oxford OX1 3QZ, UK.*

The conventional spin dynamics simulations are performed in direct products of state spaces of individual spins. In a general system of *n* spins, the total number of elements in the state basis is $\geq 4^n$. A system propagation step requires an action by an operator on the state vector and thus requires $\geq 4^{2n}$ multiplications. It's obvious that with current computers there's no way beyond about ten spins, and the calculation complexity scales exponentially with the spin system size.

We demonstrate that a polynomially scaling algorithm can be obtained if the state space is reduced by neglecting unimportant or unpopulated spin states. The class of such states is surprisingly wide. In particular, there are indications that very high multi-spin orders can be dropped completely, as can all the orders linking the spins that are remote on the interaction graph.

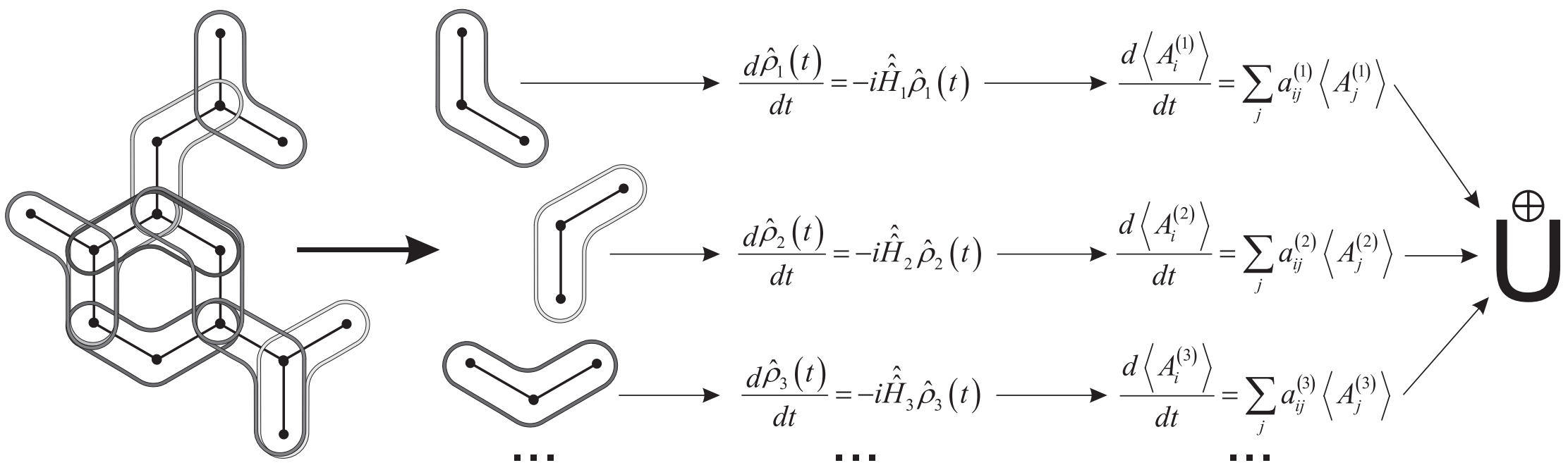

The picture above contains a schematic representation of a triples restricted calculation (straight lines denote interactions), which includes all the spin states up to the directly linked triples (e.g. $8L_Z S_\pm I_\pm$). The interaction graph is expanded into a complete set of connected subgraphs, each subgraph is treated quantum mechanically, and the resulting equations are recoupled. This procedure excludes spin orders higher than three and makes full use of the interaction topology.

The computational cost of the propagation step for a *k*tuples-restricted densely connected *n*-spin system with $k \ll n$ is $O(n^{2k})$. In cases of favourable interaction topologies (narrow graphs, e.g. in protein NMR) the asymptotic scaling is linear.